%% file: main.tex
\documentclass[submission,copyright,creativecommons]{eptcs}
\usepackage{graphicx}
\usepackage{cite}
\usepackage{glossaries}
\usepackage[colorinlistoftodos,textsize=scriptsize,textwidth=2.4cm]{todonotes}
\usepackage{xcolor}
\usepackage{tikz}
\usepackage[linesnumbered, ruled, vlined]{algorithm2e}

\graphicspath{images/}

\setacronymstyle{long-short}

\newcommand{\gwendolen}{{\sc{Gwendolen}}}

\newacronym{ai}{AI}{Artificial Intelligence}
\newacronym{bdi}{BDI}{Belief-Desire-Intention}
\newacronym{prs}{PRS}{Procedural Reasoning System}
\newacronym{qlap}{QLAP}{Qualitative Learner of Action and Perception}

\glsdisablehyper

\title{Adaptable and Verifiable BDI Reasoning\thanks{This work has been supported by the University of Liverpool's School of EEE \& CS in support of the EPSRC ``Robotics and AI for Nuclear'' (EP/R026084/1) and ``Future AI and Robotics for Space'' (EP/R026092/1) Hubs.}}

\author{
    Peter Stringer \qquad Rafael C. Cardoso \qquad Xiaowei Huang \qquad Louise A. Dennis
    \institute{University of Liverpool\\ Liverpool L69 3BX, United Kingdom}
    \email{\{peter.stringer, rafael.cardoso, xiaowei.huang, l.a.dennis\}@liverpool.ac.uk}
}
\def\titlerunning{Adaptable and Verifiable BDI Reasoning}
\def\authorrunning{P. Stringer, R.C. Cardoso, X. Huang \& L.A. Dennis}

\begin{document}
\maketitle

% Abstract

\begin{abstract}
Long-term autonomy requires autonomous systems to adapt as their capabilities no longer perform as expected. To achieve this, a system must first be capable of detecting such changes. In this position paper, we describe a system architecture for \gls{bdi} autonomous agents capable of adapting to changes in a dynamic environment and outline the required research.
Specifically, we describe an agent-maintained self-model with accompanying theories of durative actions and learning new action descriptions in \gls{bdi} systems.
%Long-term autonomy
%BDI Agent
%system ontology (self-model)
%Verification

\end{abstract}

\section{Introduction}

Long-term autonomy requires autonomous systems to adapt as their capabilities no longer perform as expected. To achieve this, a system must first be capable of detecting such changes. Creating and maintaining a \emph{system ontology} is a comprehensive solution for this; an agent-maintained formal \emph{self-model} will take the role of this system ontology. It would act as a repository of information about all the processes and functionality of the autonomous system, forming a systematic approach for detecting action failures.

Our work will focus on \acrfull{bdi}\cite{rao1992abstract} programming languages as they are well known for their use in developing intelligent agents \cite{aitken17:_auton_nuclear_waste_manag, dennis2010satellite, harland2014operational, mascardi2005languages}. Agents that are capable of controlling an array of cyber-physical autonomous systems such as autonomous vehicles, spacecraft and robot arms have been programmed using \gls{bdi} agents (e.g., Mars Rover\cite{harland2014operational}, Earth-orbiting satellites\cite{dennis2010satellite} and robotic arms for nuclear waste-processing\cite{aitken17:_auton_nuclear_waste_manag}). Coupled with their use of plans and actions, \gls{bdi} languages offer an appropriate platform to build upon for the development of an adaptable autonomous system.

The agent-maintained self-model includes \emph{action descriptions}, consisting of pre- and post-conditions of all known actions/capabilities. An action's pre-conditions are the environment conditions that must exist for an action to be executed whilst post-conditions are defined as the expected changes in the environment made directly by a completed action. These action descriptions are based on the Planning Domain Definition Language (PDDL)~\cite{Mcdermott98}, commonly used in classical automated planning. The complete availability of current system information will provide the ability to monitor the status of actions, presenting the opportunity to detect failure. We use \emph{action life-cycles} based on a theory of durative actions for \gls{bdi} systems~\cite{dennis2014actions} to detect persistent abnormal behaviour from action executions that could denote hardware degradation or other long-term causes of failure such as exposure to radiation or extreme temperature. Once a failure has been detected, we can use machine learning methods to update the action description in the self model. Then, we can repair or replace the actions in any existing plans by using an automated planner to patch these plans. The resulting plans can then be verified to ensure the system’s safety properties are intact.

This is a position paper that outlines a program of research. The overarching aim of this research is to create a framework for the verification of autonomous systems that are capable of learning new behaviour(s) descriptions and integrating them into existing \gls{bdi} plans: using the framework as a route to certification. In this paper, we discuss the current ability of \gls{bdi} systems in adaptable reasoning, largely focusing on actions. We also consider research in \gls{ai} planning on modelling actions and the methods and implications of introducing machine learning for replacing action descriptions. Our main contribution is the initial design of a system architecture for \gls{bdi} autonomous agents capable of adapting to changes in a dynamic environment, consolidating the agent-maintained self-model with the theory of durative actions and learning new action descriptions into a cohesive and adaptable \gls{bdi} system. It should be noted, our work relies upon assumptions that have been discussed further in the relevant sections of this paper.
%\ps{The reviewers comment stated that contributions and assumptions should be discussed together... I can't find a way of mentioning the very technical assumptions so early on in the paper, although I also can't remove the contributions from the introduction}
%\rcnote{A possible solution would be to enumerate the assumptions and make them bold font. However, I don't think this is needed in a short paper. Could be useful for the thesis.}
%\section{Background}

\section{The Belief-Desire-Intention Model of Agency}
Intelligent Agent systems conforming to the \gls{bdi} software model largely follow the principles proposed by Bratman's \textit{Intentions, Plans, and Practical Reason}\cite{bratman1987intention} that was originally intended for modelling practical reasoning in human psychology. Use of \gls{bdi} agents is particularly suitable for high-level management and control tasks in complex dynamic environments\cite{rao1995bdi} which justifies their implementation in many practical applications. Since Georgeff and Lansky's \gls{prs} emerged in 1987 \cite{georgeff1987reactive} a wide range of \gls{bdi} programming languages have been developed \cite{mascardi2005languages}, each building upon the reach of \gls{prs} with a multitude of extensions for different applications.

\subsection{Actions in BDI Systems}
%Rafael: Commented that sentence because I think it needs a citation and more discussion around it.
%BDI languages have been described as reactive planning languages.  
Typically, BDI languages use \emph{plans} provided by a programmer at compile time and the language selects an appropriate plan to react to a given situation. Some BDI languages model interaction with the external environment either as an \emph{action} (e.g., Jason~\cite{Bordini07})
or as a \emph{capability} (e.g., GOAL~\cite{Hindriks00}).  We view capabilities as actions with explicit pre- and post-conditions.  

Actions and capabilities can appear in the bodies of \emph{plans}.  The body of a plan is generally a sequence of actions/capabilities, belief updates and subgoal manipulations (e.g., adopting or dropping goals).  Plans are selected by means-end reasoning in order to achieve the agent's goals.  Plans may have additional components as well as the plan body. For instance, they generally have a \emph{guard} which must hold before the plan can be applied. Once a plan is selected for execution it is transformed into an \emph{intention} which represents a sequence of steps to be performed as part of executing the plan.

We intend to extend the \gwendolen\ agent programming language\cite{dennis17gwen} for our research. We have chosen to use \gwendolen\ as it is a \gls{bdi} agent programming language capable of producing verifiable agents. It is integrated into the MCAPL (Model-Checking Agent Programming Languages) framework~\cite{dennis2018mcapl}. Using MCAPL, agents can be programmed in \gwendolen\ and then verified using the AJPF (Agent Java Pathfinder) model-checker~\cite{Dennis2012}. Actions in \gwendolen\ are generally implemented in a Java-based environment; at runtime they are requested and executed by agents. Whilst actions in \gwendolen\ can exhibit characteristics of duration, this is implemented using a `wait for' construct which temporarily suspends an intention when encountered. When the predicate that is being waited for is believed, the intention becomes unsuspended. This is the extent to which actions in \gwendolen\ are treated as having significant durations and is largely typical of the treatment of actions in \gls{bdi} languages.

\subsection{Durative Actions}
\gls{bdi} languages are increasingly being used for developing agents for physical systems where actions could take considerable time to complete \cite{dennis2014actions}. Currently, most BDI languages suspend an agent entirely until an action completes or implement actions in such a way that an agent may start a process but then must be programmed to explicitly track the progress of the action in some way. 

Introducing an explicit notion of duration to actions will allow us to create principled mechanisms to let an agent continue operating once an action is started, meaning the agent is available to monitor the status of actions in progress. \cite{harland2014operational} introduced an abstract theory of \emph{goal life-cycles}, whereby every goal pursued by the agent moves through a series of states: \emph{Pending} to \emph{Active}; \emph{Active} to either \emph{Suspended} or \emph{Aborted} or a \emph{Successful} end state; and so on. Dennis and Fisher \cite{dennis2014actions} extended the formal semantics provided by Harland et al. to show how the behaviour of durative actions could integrate into these life-cycles. They advocate associating actions not only with pre- and post-conditions containing durations but also with explicit success, failure and abort conditions (an abort is used if the action is ongoing but needs to be stopped) and suggest goals be suspended while an action is executing and then the action's behaviour be monitored for the occurrence of its success, failure or abort conditions.  When one of these occurs the goal then moves to the \emph{Active} or \emph{Pending} (where re-planning may be required) part of its life-cycle as appropriate. Adding these additional states to actions should not add to the cost of model checking as this should not add branches. Adding states should only add more information which would make no significant difference.

Brahms \cite{sierhuis2007brahms}, a multi-agent modelling environment, is an example of an agent approach that implements durative actions. The Brahms equivalent of actions, \emph{activities}, have duration.  Brahms has a formal semantics provided by Stocker et al. \cite{stocker2011formal}, although these semantics are primarily concerned with the effect of activity duration on simulation with mechanism for monitoring the behaviour of an activity during its execution. Whilst the concept of durative actions seems to have been adequately explored in these examples, there has not been a formal implementation that focuses on monitoring individual actions for failure.

\subsection{Action Failure}

The idea of monitoring an action’s life-cycle exists in current literature~\cite{harland2017aborting, harland2014operational, dennis2014actions}. A range of states can be attributed to an action that can subsequently be traced for irregularities or consistent errors, providing a basis for determining failure. If we assume that the performance of actions may degrade then we also need to introduce the concept of an action life-cycle in which an action is introduced into the system as \emph{Functional}, may move into a \emph{Suspect} state if it is failing and finally becomes \emph{Deprecated} following repeated failures.

% has been introduced, actions can be monitored throughout their life-cycle. This is critical for detecting abnormality and consequent failure. Similar monitoring techniques have been applied to \gls{bdi} agent goals, with goals residing in one of the following states at any time: Pending; Active; Monitoring; Suspended; and Aborting \cite{harland2014operational,harland2017aborting}. Maintaining a repository of these states for actions too would greatly extend the ability of a BDI agent.\LD{This feels like repetition rather than expansion and clarification - I think maybe introducing the diagram of the action lifecycle and explaining how action descriptions should be extended to include success/failure etc. states and then execution would automatically track these.}

% \subsubsection{Action Failure}
Cardoso et al. \cite{cardosoplan} assumes a framework along these lines and builds upon it to outline a mechanism that allows reconfiguration of the agent's plans in order to continue functioning as intended if some action has become \emph{Deprecated}. However, this assumed ability to detect persistent failures does not yet exist. Our proposed framework should allow us to detect persistent abnormal behaviour from action executions for use with Cardoso et al.'s reconfiguration mechanism.

%Cardoso et al. used the Mars \emph{Curiosity} Rover as a practical example for their work on plan library reconfigurability. In their scenario the rover developed a faulty movement capability, requiring intervention. A fault such as this could be detected by comparing the original action's post-conditions with the perceived post-conditions detected after execution.

%\LD{yes, you definitely need more above if you are going to mention post-conditions here/}

%^^^\LD{Maybe move this earlier to life-cycles?}

\section{AI Planners and Learning New Actions}

\gls{ai} Planning seeks to automate reasoning about plans; using a formal description of the domain, all possible actions available in the domain, an initial state of the problem, and a goal condition to produce a plan consisting of the actions that will achieve the goal condition when executed \cite{haslum2019introduction}. The formal description of the domain and the problem can be considered a model of the environment, the accuracy of which is fundamental to producing viable plans of reasonable quality. Significant advances have been made in the modelling of actions \cite{fox2003pddl2, ghallab2016automated, weld2008planning, younes2004solving} in automated planning, supporting actions that can have variable duration, conditions and effects.

%\subsection{Learning New Actions}
Actions in \gls{bdi} systems are typically designed without a specified duration and are defined before the execution of the program. As previously mentioned, \gls{bdi} systems do not have a de facto theory of durative actions. Additionally, there is no theory for learning new action descriptions. With an extension of action theory for \gls{bdi} systems (covering these two areas) paired with the self-model concept, actions could adapt to change. However, learning a new action may not always be the best solution for a failing action. Cardoso et al. \cite{cardosoplan} have developed a method for reasoning about replacing malfunctioning actions with alternate existing actions to achieve the same desired goal, reusing the domain entities and predicates that are already available. 

In situations where a new action description is required at runtime, there are already suitable learning methods that could be adapted to be incorporated into the framework \cite{mugan2011autonomous , pasula2007learning}, enabling the discovery of new entities and predicates in the domain. In \cite{mugan2011autonomous}, the \gls{qlap} is introduced. When deployed in an unknown, continuous and dynamic environment, \gls{qlap} constructs a hierarchical structure of possible actions in an environment based upon the consequences of actions that have happened before. Work in \cite{pasula2007learning} explores the use of Machine Learning and probabilistic planning in complex environments to cope with unexpected outcomes. A learning algorithm is used to determine an action model with the greatest likelihood of attaining the perceived action effects of another different set of actions. We have to acknowledge the risk that system properties could be violated during machine learning, although this could be remedied by using a \emph{Safe Learning}\cite{JMLR:v16:garcia15a} approach, the introduction of machine-learning presents great difficulty for verification as the algorithms cannot (currently) be directly verified~\cite{van2017challenges}. As a consequence, it should be noted that the proposed system could be unsuitable for scenarios where learning from failure is not safe (e.g., autonomous drones), where it would be safest to execute a controlled stop to the system rather than attempting a recovery.

% How should I go about answering the feedback controller question
%In some scenarios, a feedback controller would suffice for correcting failure. However, numerous examples exist where a software solution would exceed the capabilities of such hardware. Ultimately, unless every mechanical component on the system has been coupled with hardware for failure recovery, the system is still susceptible to total failure. For remotely deployed systems such as the Mars Curiosity, this could result in mission failure and astronomical financial loss.\LD{I'm not sure I really understand this paragraph}

%\section{Self-Model}

\section{System Architecture}

\begin{figure}[ht]
\centering{
%\resizebox{125mm}{!}{\input{SystemDiagramAREA2020.pdf_tex}}
\fontsize{8pt}{8pt}\selectfont
\def\svgwidth{\linewidth}
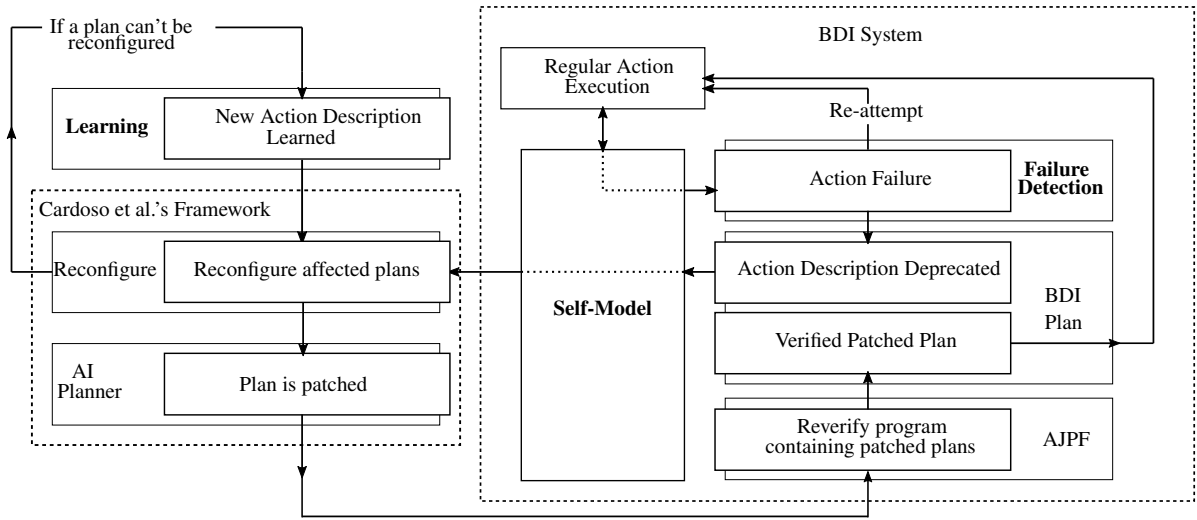
\caption{Diagram of Action Failure and Recovery Mechanisms. Arrows represent data flow and dotted lines are for readability when a line goes through a component.}
\label{fig:topView}
}
\end{figure}

The initial objective of this research is to formally define the concept of a self-model: an agent-maintained ontology for the autonomous system. We intend to use PDDL (Planning Domain Definition Language) as a starting point for creating a self-model. PDDL is a formalism for \gls{ai} planning which is intended to ``express the `physics' of a domain'' ~\cite{Mcdermott98}. %It shares a similar action theory with pre- and post- conditions for actions with some \gls{bdi} languages. 
More specifically, we intend to use features introduced in PDDL2.1 \cite{fox2003pddl2} as our starting point. PDDL2.1 is an extension to PDDL for expressing temporal planning domains. The self-model concept will build on this by enabling agents to access and maintain a domain description, adding the capability of learning new action descriptions and allowing action life-cycles to be monitored. As shown in Figure~\ref{fig:topView}, the self-model is centrally linked to the other system components as they are required to contribute into keeping the self-model accurate. It is important to note that the self-model's domain description is not assumed to be modelled soundly and completely, yet it is assumed that all reports and updates received by the system are correct.
%\rcnote{I am not sure about this. It is fine for this paper but we have to discuss it in a meeting.}

Our implementation will be developed for \gwendolen~\cite{dennis2008gwendolen}. The \gwendolen\ agent programming language follows the \gls{bdi} software model. As part of the MCAPL Framework, \gwendolen\ interfaces with the Java Pathfinder (JPF) model-checker\cite{Visser03}. Our intention is to implement self-models and the theory of action life-cycles~\cite{dennis2014actions} in \gwendolen\ and integrate this with the existing work on plan reconfigurability~\cite{cardosoplan}.  We will then exploit \gwendolen's support for verification to verify the adapted system against requirements. We propose representing actions in our self-model with explicit pre- and post-conditions and either explicit success, fail and abort conditions or one's that can be inferred from the pre- and post-conditions. We will then adapt the \gwendolen\ goal life-cycle as suggested in~\cite{dennis2014actions} to handle durative actions in a principled fashion. 

%\gwendolen will be used to showcase the \textit{Self-Model} as there is already scope for durative actions and an existing model-checking capability in place for verifying agent programs that could be used to verify the patched plans containing adapted actions that have been learned using Machine Learning.

%However, there are solutions in development for verifying the results of machine-learning algorithms~\cite{huang2017safety} that will be considered in future work.\LD{I think this is a red herring}

\begin{algorithm}
\caption{Action Failure Monitoring\label{AFR}}
 \SetKwFunction{FMain}{monitor}
 \SetKwProg{Fn}{Function}{}{}
 \Fn{\FMain{$\langle$action\_identifier, action\_status, action\_post\_conditions$\rangle$}}{
$ActionID \longleftarrow action\_identifier$\;
$Status  \longleftarrow action\_status$\;
\If{$Status \neq Active \lor Failed$}{\Return $Status$\;}
$ExpectedPostCond  \longleftarrow action\_post\_conditions$\;
\While{Status = Active}{
$ActualPostCond  \longleftarrow$ getPostConditions(ActionID)\;
\uIf{ActualPostCond = ExpectedPostCond}{
$Status  \longleftarrow Active$\;
monitor($ActionID$, $Status$, $ActualPostCond$)\;
}\Else{
$Status  \longleftarrow Failed$\;
reconfigure($ActionID$)\;
\Return $Status$\;
}
}
\Return $Active$\;
}
\end{algorithm}

When an action changes, requiring plans to be modified, it is assumed that the agent must be verified again in order to preserve the safety properties of the system as a whole. However, if a new action is learnt in place of a failing action (fully or partially achieving the failing action’s post-conditions), the whole system may not require reverification. We aim to further study this process in order to identify the conditions where such reverification would not be necessary.

%This computationally expensive task could be avoided by only reverifying the components that have been affected by the new action which could prove crucial in an environment where resources (such as battery power and time) are limited. Therefore, a successful implementation of this could have numerous practical applications considering the benefits of conservative power usage and increased efficiency.\LD{I think drop or comment out this paragraph for now -- quite how compositional verification might work in this framework is begging a lot of questions}

In \emph{Algorithm 1}, we propose a primitive method for action failure monitoring. It is assumed that the action status used in the algorithm is asserted as a belief by the system. We start monitoring an action once it has been executed, retrieving some preliminary information about the action: an identifier and the current status (lines 2-3). If the action is currently \emph{Pending}; \emph{Suspended}; or \emph{Aborting}, this status is returned (lines 4-5). If not, the action's expected post-conditions are retrieved from the tuple (line 6). Whilst an action's state is '\emph{Active}', we continue checking it for failure by comparing the perceived post-conditions with those that are expected of that action (lines 7-11). If at any point during monitoring these conditions do not match, the action's state becomes '\emph{Failed}'. If an action is not working as expected, the action can be re-attempted or suspended and replaced in the self-model. The replacement action may be selected from an existing action in the self-model itself. Alternatively, using machine-learning methods, a new action can be learnt to replace the failing action using current knowledge of the available capabilities. Finally, a method for reconfiguring the BDI plan, such as in \cite{cardosoplan}, is called (lines 12-15).

\subsection{Scenario}
To illustrate how the self-model would complement Cardoso et al.'s work on reconfigurability \cite{cardosoplan}, we use the same scenario: a Mars rover's faulty movement capability. Figure~\ref{fig:topView} shows our proposed mechanisms for actions failure and recovery embedded into a system architecture including a BDI system, an \gls{ai} Planner and Cardoso et al.'s \cite{cardosoplan} reconfigurability framework. The dotted line arrows crossing the self-model represent incoming information from a component such as an action's state. The system architecture in the diagram relies upon a simplification of the successful, fault-free execution of actions that would normally occur in a BDI system. In the case of the rover, these fault-free actions can be represented by the high-level task of movement between waypoints. Whilst mostly successful, these actions are susceptible to failure.

Consider the task of moving from a waypoint A to another waypoint B, in order to collect a rock sample to analyse at another waypoint C. Using the monitoring method for failure detection in Algorithm~\ref{AFR}, a failed movement action between any of these waypoints could be found. Given the dynamic environment that the rover operates in, it is plausible that previously clear and usable routes could become blocked at any time. A failure can be flagged when an action is exceeding a predetermined time or energy threshold described in the action post-conditions. Once failure has been detected and confirmed, we can update the self-model to show that the action description has deprecated and no longer affords it's post-conditions. The rover now attempts to reconfigure the current plan to resolve the failure using an \gls{ai} planner to search for a replacement (e.g., by finding a different route) before attempting to learn a completely new action description. In both cases, the time and energy consumption required to accomplish the original post-conditions is updated in the reconfigured/new action description. If it is found that the reconfigured plan is now too time or energy intensive, the latter method of learning a new action description is invoked. If at any point the failing action is found to achieve all post-conditions but does not perform the action within the time or energy threshold (e.g., the rover now navigates around a blockage and arrives at the correct waypoint but now takes longer to do so), this can be managed by learning new actions descriptions with an updated time and/or energy threshold. 

The action may not be deprecated if the failure is considered anomalous; for instance, the action normally succeeds and only fails on one isolated occasion.  If the action description is not deprecated, then the action will be re-attempted without resorting to reconfiguration or learning methods. The rover can continue progressing towards the goal if the failure was anomalous. If a new action description is learned, the original plan will be patched with the new action description by the \gls{ai} planner. This plan could require reverification to preserve previously verified properties which can be handled by using AJPF. Ensuring these properties are maintained is crucial for avoiding failure. The verified patched plan can then be used in the BDI system where regular action execution continues and the rover can continue to complete it's mission.

\section{Related Work}

The work in~\cite{cardosoplan} describes a reconfigurability framework that is capable of replacing faulty action descriptions based on formal definitions of action descriptions, plans, and plan replacement. The implementation uses an \gls{ai} planner to search for viable action replacements. We plan on extending their approach by adding the concept of a self-model, durative actions, and failure detection. Furthermore, we also envision adding a learning component to the framework in order to be able to cope with dynamic environment events that require new action descriptions to be formulated at runtime.

Troquard et al.'s work on logic for agency in \cite{troquard2006towards} considers the modelling of actions with durations although a different approach was taken: actions are given duration using continuations from STIT (Seeing To It That) logic. In \gls{bdi} systems, the focus of handling plan failure is the effect that failure has on goals \cite{bordini2010semantics, sardina2011bdi}. This is a reasonable focus considering the central role that goals have in agent-oriented programming. Consequently, action failure recovery has not been explored as an option for managing plan failure.

\section{Conclusions and Future Work}

In this position paper we have described a system architecture for \gls{bdi} autonomous agents capable of adapting to changes in a dynamic environment. We also introduced the idea of an agent-maintained self-model with durative actions and learning new action descriptions. Our proposed system aims to resolve the following: develop the concept of a self-model; produce and develop a method to detect the failure of an action performed by a \gls{bdi} Agent; develop a theory of durative actions for \gls{bdi} languages; adapt existing system to allow new actions to be learnt and used in place of failing ones whilst preserving safety properties, and finally to integrate into the existing \gwendolen\ infrastructure.

To illustrate the applicability of the discussed mechanisms, a practical example of how a Mars rover could make use of the framework was provided. Future work includes defining the learning component to be able to handle dynamic environment events that require the creation of new action descriptions at runtime, a formal definition of the self-model with an outline of the concepts included in this, the implementation of the system architecture, and the evaluation of the approach.

A number of questions and challenges have been identified whilst outlining this program of research. Firstly, it has been noted that the term 'persistent failure' is subjective and should be accompanied by a formal and precise specification to avoid ambiguity. Secondly, considerations for the steps taken after reconfiguration and the learning process require further work (e.g. What happens to failing actions in the model after reconfiguring?). Finally, the proposed learning strategy has produced many challenges which will be considered once implementation has reached this stage. Notably, we will consider how the learning method can ensure valid solutions; how planning time could be minimised and how an action's state could influence the learning strategy.
These challenges will serve as guidance for future work.

\nocite{*}
\bibliographystyle{eptcs}
\bibliography{bibliography}
\end{document}

%% file: SystemDiagramAREA2020.pdf_tex
%% Accompanies image file 'SystemDiagramAREA2020.pdf' (pdf, eps, ps)
\begingroup%
  \makeatletter%
  \providecommand\color[2][]{%
    \errmessage{(Inkscape) Color is used for the text in Inkscape, but the package 'color.sty' is not loaded}%
    \renewcommand\color[2][]{}%
  }%
  \providecommand\transparent[1]{%
    \errmessage{(Inkscape) Transparency is used (non-zero) for the text in Inkscape, but the package 'transparent.sty' is not loaded}%
    \renewcommand\transparent[1]{}%
  }%
  \providecommand\rotatebox[2]{#2}%
  \newcommand*\fsize{\dimexpr\f@size pt\relax}%
  \newcommand*\lineheight[1]{\fontsize{\fsize}{#1\fsize}\selectfont}%
  \ifx\svgwidth\undefined%
    \setlength{\unitlength}{885.3246013bp}%
    \ifx\svgscale\undefined%
      \relax%
    \else%
      \setlength{\unitlength}{\unitlength * \real{\svgscale}}%
    \fi%
  \else%
    \setlength{\unitlength}{\svgwidth}%
  \fi%
  \global\let\svgwidth\undefined%
  \global\let\svgscale\undefined%
  \makeatother%
  \begin{picture}(1,0.43795628)%
    \lineheight{1}%
    \setlength\tabcolsep{0pt}%
    \put(0,0){\includegraphics[width=\unitlength,page=1]{SystemDiagramAREA2020.pdf}}%
    \put(0.0549356,0.32634866){\makebox(0,0)[lt]{\lineheight{1.25}\smash{\begin{tabular}[t]{l}\textbf{Learning}\\\end{tabular}}}}%
    \put(0,0){\includegraphics[width=\unitlength,page=2]{SystemDiagramAREA2020.pdf}}%
    \put(0.67318411,0.28276066){\color[rgb]{0,0,0}\makebox(0,0)[lt]{\lineheight{1.25}\smash{\begin{tabular}[t]{l}Action Failure\end{tabular}}}}%
    \put(0,0){\includegraphics[width=\unitlength,page=3]{SystemDiagramAREA2020.pdf}}%
    \put(0.46047285,0.17440098){\makebox(0,0)[lt]{\lineheight{1.25}\smash{\begin{tabular}[t]{l}\textbf{Self-Model}\end{tabular}}}}%
    \put(0,0){\includegraphics[width=\unitlength,page=4]{SystemDiagramAREA2020.pdf}}%
    \put(0.19961011,0.1115912){\makebox(0,0)[lt]{\lineheight{1.25}\smash{\begin{tabular}[t]{l}Plan is patched\end{tabular}}}}%
    \put(0,0){\includegraphics[width=\unitlength,page=5]{SystemDiagramAREA2020.pdf}}%
    \put(0.0601783,0.12259798){\makebox(0,0)[lt]{\lineheight{1.25}\smash{\begin{tabular}[t]{l}   AI\\\\\end{tabular}}}}%
    \put(0,0){\includegraphics[width=\unitlength,page=6]{SystemDiagramAREA2020.pdf}}%
    \put(0.45305262,0.37566693){\color[rgb]{0,0,0}\makebox(0,0)[lt]{\lineheight{1.25}\smash{\begin{tabular}[t]{l}Regular Action\\\end{tabular}}}}%
    \put(0.86872113,0.18535306){\makebox(0,0)[lt]{\lineheight{1.25}\smash{\begin{tabular}[t]{l}BDI\\Plan\\\end{tabular}}}}%
    \put(0.61350164,0.20807077){\color[rgb]{0,0,0}\makebox(0,0)[lt]{\lineheight{1.25}\smash{\begin{tabular}[t]{l}Action Description Deprecated\end{tabular}}}}%
    \put(0.86721285,0.06609341){\makebox(0,0)[lt]{\lineheight{1.25}\smash{\begin{tabular}[t]{l}AJPF\\\end{tabular}}}}%
    \put(0.84732238,0.29044725){\makebox(0,0)[lt]{\lineheight{1.25}\smash{\begin{tabular}[t]{l}\textbf{  Failure }\\\end{tabular}}}}%
    \put(0,0){\includegraphics[width=\unitlength,page=7]{SystemDiagramAREA2020.pdf}}%
    \put(0.1795768,0.33453592){\color[rgb]{0,0,0}\makebox(0,0)[lt]{\lineheight{1.25}\smash{\begin{tabular}[t]{l}New Action Description \end{tabular}}}}%
    \put(0,0){\includegraphics[width=\unitlength,page=8]{SystemDiagramAREA2020.pdf}}%
    \put(0.04514376,0.20712854){\makebox(0,0)[lt]{\lineheight{1.25}\smash{\begin{tabular}[t]{l}Reconfigure\\\end{tabular}}}}%
    \put(0,0){\includegraphics[width=\unitlength,page=9]{SystemDiagramAREA2020.pdf}}%
    \put(0.16165085,0.20678557){\color[rgb]{0,0,0}\makebox(0,0)[lt]{\lineheight{1.25}\smash{\begin{tabular}[t]{l}Reconfigure affected plans\end{tabular}}}}%
    \put(0,0){\includegraphics[width=\unitlength,page=10]{SystemDiagramAREA2020.pdf}}%
    \put(0.66289151,0.07632297){\makebox(0,0)[lt]{\lineheight{1.25}\smash{\begin{tabular}[t]{l}Reverify program \\\end{tabular}}}}%
    \put(0.64472555,0.14822501){\color[rgb]{0,0,0}\makebox(0,0)[lt]{\lineheight{1.25}\smash{\begin{tabular}[t]{l}Verified Patched Plan\end{tabular}}}}%
    \put(0,0){\includegraphics[width=\unitlength,page=11]{SystemDiagramAREA2020.pdf}}%
    \put(0.04191921,0.40954208){\color[rgb]{0,0,0}\makebox(0,0)[lt]{\lineheight{1.25}\smash{\begin{tabular}[t]{l}If a plan can't be \end{tabular}}}}%
    \put(0,0){\includegraphics[width=\unitlength,page=12]{SystemDiagramAREA2020.pdf}}%
    \put(0.68083404,0.40218787){\color[rgb]{0,0,0}\makebox(0,0)[lt]{\lineheight{1.25}\smash{\begin{tabular}[t]{l}BDI System\end{tabular}}}}%
    \put(0,0){\includegraphics[width=\unitlength,page=13]{SystemDiagramAREA2020.pdf}}%
    \put(0.46986034,0.35946197){\color[rgb]{0,0,0}\makebox(0,0)[lt]{\lineheight{1.25}\smash{\begin{tabular}[t]{l}Execution\end{tabular}}}}%
    \put(0.8463137,0.27514625){\makebox(0,0)[lt]{\lineheight{1.25}\smash{\begin{tabular}[t]{l}\textbf{Detection}\\\end{tabular}}}}%
    \put(0.6897818,0.33955326){\color[rgb]{0,0,0}\makebox(0,0)[lt]{\lineheight{1.25}\smash{\begin{tabular}[t]{l}Re-attempt\end{tabular}}}}%
    \put(0,0){\includegraphics[width=\unitlength,page=14]{SystemDiagramAREA2020.pdf}}%
    \put(0.22206778,0.31708423){\color[rgb]{0,0,0}\makebox(0,0)[lt]{\lineheight{1.25}\smash{\begin{tabular}[t]{l}Learned\end{tabular}}}}%
    \put(0,0){\includegraphics[width=\unitlength,page=15]{SystemDiagramAREA2020.pdf}}%
    \put(0.0332205,0.25704364){\color[rgb]{0,0,0}\makebox(0,0)[lt]{\lineheight{1.25}\smash{\begin{tabular}[t]{l}Cardoso et al.'s Framework\end{tabular}}}}%
    \put(0.05738174,0.39603162){\color[rgb]{0,0,0}\makebox(0,0)[lt]{\lineheight{1.25}\smash{\begin{tabular}[t]{l}reconfigured\end{tabular}}}}%
    \put(0.04831824,0.10565505){\makebox(0,0)[lt]{\lineheight{1.25}\smash{\begin{tabular}[t]{l}  Planner\\\\\end{tabular}}}}%
    \put(0.63674558,0.06009162){\makebox(0,0)[lt]{\lineheight{1.25}\smash{\begin{tabular}[t]{l}containing patched plans\\\end{tabular}}}}%
  \end{picture}%
\endgroup%